\documentclass[%
reprint,
superscriptaddress,
 amsmath,amssymb,
 aps,
prb,
]{revtex4-2}

\usepackage{graphicx}
\usepackage{dcolumn}
\usepackage{ulem}
\usepackage{bm}
\usepackage[colorlinks=true,linkcolor=blue,citecolor=blue]{hyperref}
\usepackage{siunitx}
\usepackage{xcolor}
\usepackage{natbib}
\usepackage{gensymb}
\usepackage{parskip}
\begin{document}

\title{Recycling of Perovskite Substrate}

\author{Jie Wang}
\thanks{These authors contributed equally.}
\affiliation{Faculty of Materials Science and Engineering, Kunming University of Science and Technology, Kunming, 650093, Yunnan, China}

\author{Yuhan Liang}
\thanks{These authors contributed equally.}
\affiliation{School of Materials Science and Engineering, Tsinghua University, Beijing, 100084, China}

\author{Yue Wang}
\affiliation{School of Materials Science and Engineering, Tsinghua University, Beijing, 100084, China}

\author{Shengan Yang}
\affiliation{Faculty of Materials Science and Engineering, Kunming University of Science and Technology, Kunming, 650093, Yunnan, China}

\author{Xinyu Zhang}
\affiliation{Faculty of Materials Science and Engineering, Kunming University of Science and Technology, Kunming, 650093, Yunnan, China}

\author{Xin Huang}
\affiliation{Faculty of Materials Science and Engineering, Kunming University of Science and Technology, Kunming, 650093, Yunnan, China}

\author{Yajuan Wang}
\affiliation{Faculty of Materials Science and Engineering, Kunming University of Science and Technology, Kunming, 650093, Yunnan, China}

\author{Zhaowei Liang}
\affiliation{Faculty of Materials Science and Engineering, Kunming University of Science and Technology, Kunming, 650093, Yunnan, China}

\author{Ji Ma}
\affiliation{Faculty of Materials Science and Engineering, Kunming University of Science and Technology, Kunming, 650093, Yunnan, China}

\author{Hui Zhang}
\affiliation{Faculty of Materials Science and Engineering, Kunming University of Science and Technology, Kunming, 650093, Yunnan, China}

\author{Qingming Chen}
\affiliation{Faculty of Materials Science and Engineering, Kunming University of Science and Technology, Kunming, 650093, Yunnan, China}

\author{Jing Ma}
\affiliation{School of Materials Science and Engineering, Tsinghua University, Beijing, 100084, China}

\author{Yuanhua Lin}
\email{linyh@tsinghua.edu.cn}
\affiliation{School of Materials Science and Engineering, Tsinghua University, Beijing, 100084, China}

\author{Liang Wu}
\email{liangwu@kust.edu.cn}
\affiliation{Faculty of Materials Science and Engineering, Kunming University of Science and Technology, Kunming, 650093, Yunnan, China}

\date{\today}

\begin{abstract}

The use of water-soluble sacrificial layer of Sr$_3$Al$_2$O$_6$ has tremendously boosted the research on freestanding functional oxide thin films, especially thanks to its ultimate capability to produce high-quality epitaxial perovskite thin films. However, the costly single-crystalline substrates, e.g. SrTiO$_3$, were generally discarded after obtaining the freestanding thin films. Here, we demonstrate that the SrTiO$_3$ substrates can be recycled to fabricate La$_{0.7}$Sr$_{0.3}$MnO$_3$ films with nearly identical structural and electrical properties. After attaining freestanding thin films, the residues on SrTiO$_3$ can be removed by 80 \degree C hot water soaking and rinsing treatments. Consequently, the surface of SrTiO$_3$ reverted to its original step-and-terrace structure.

\end{abstract}

\maketitle

\section{Introduction}

The freestanding complex oxides open a door to additional spectrums of two-dimensional materials with exotic correlated electronic phases. Based on the excellent stretchability of freestanding oxide membranes and strong correlation between quantum properties and lattice constant, the functional properties of oxide can be comparable and even superior to those in their bulk or rigid thin film forms, such as enhanced and controllable ferromagnetism \cite{Lu2016, Hong2020, Peng2022}, monolayer and super-elastic ferroelectricity \cite{Ji2019,Dong2019}, and topological ferroelectric domains \cite{Han2022}. Most recently, the freestanding complex oxides were also demonstrated to integrate with conventional two-dimensional transition metal dichalcogenides as a high $\kappa$ layer \cite{Yang2022,Huang2022}. Thus, the freestanding complex oxide provides a promising path for approaching applicable flexible devices with desirable quantum properties. This thriving area has been already well covered in several review articles, see for example Refs. \cite{Zhang2019,Kum2019,Gao2020}.

Among the various epitaxial lift-off techniques to fabricate freestanding complex oxides, the water-soluble sacrificial Sr$_3$Al$_2$O$_6$ (SAO) buffer layer exhibits many advantages in the epitaxy growth of high-quality transition metal perovskites on high-priced parent single-crystalline substrates. In the related fields, such as semiconductor, solar cell, and 2D material fabrications, wafer and substrate recycling is of great importance in reducing the cost \cite{Lee2014, Wie2018, Huang2016, Huang2017, Takada2020, Xu2015, Wang2017}. However, these substrates, such as SrTiO$_3$, were generally discarded. To this end, to develop a recycling procedure is important.

Here, we demonstrate that a La$_{0.7}$Sr$_{0.3}$MnO$_3$ (LSMO) grown on a recycled SrTiO$_3$ (STO) substrate displays practically identical structural and transport properties. Here, the debris on the substrate after room-temperature water soaking (to get freestanding thin films) can be completely removed by adequate hot water (80 \degree C) soaking and rinsing, and the perovskite substrates are recovered to their original step-and-terrace surface structure with indiscernible debris by atomic force microscopy (AFM), or any residual Al$^{3+}$ from adjacent Sr$_3$Al$_2$O$_6$ layer by X-ray photoelectron spectroscopy (XPS). The hot water treatment could also be conducive to obtaining residue-free freestanding films, especially when the interface is extremely vital in freestanding multi-layer devices. It is worth noting that, such substrate recycling can be applied not only to prepare freestanding thin films but also to synthesize epitaxial functional thin films and devices by artificially introducing the SAO as an excellent degradable buffer layer.  

\begin{figure}[ht]
\centering
	\includegraphics[width=\columnwidth]{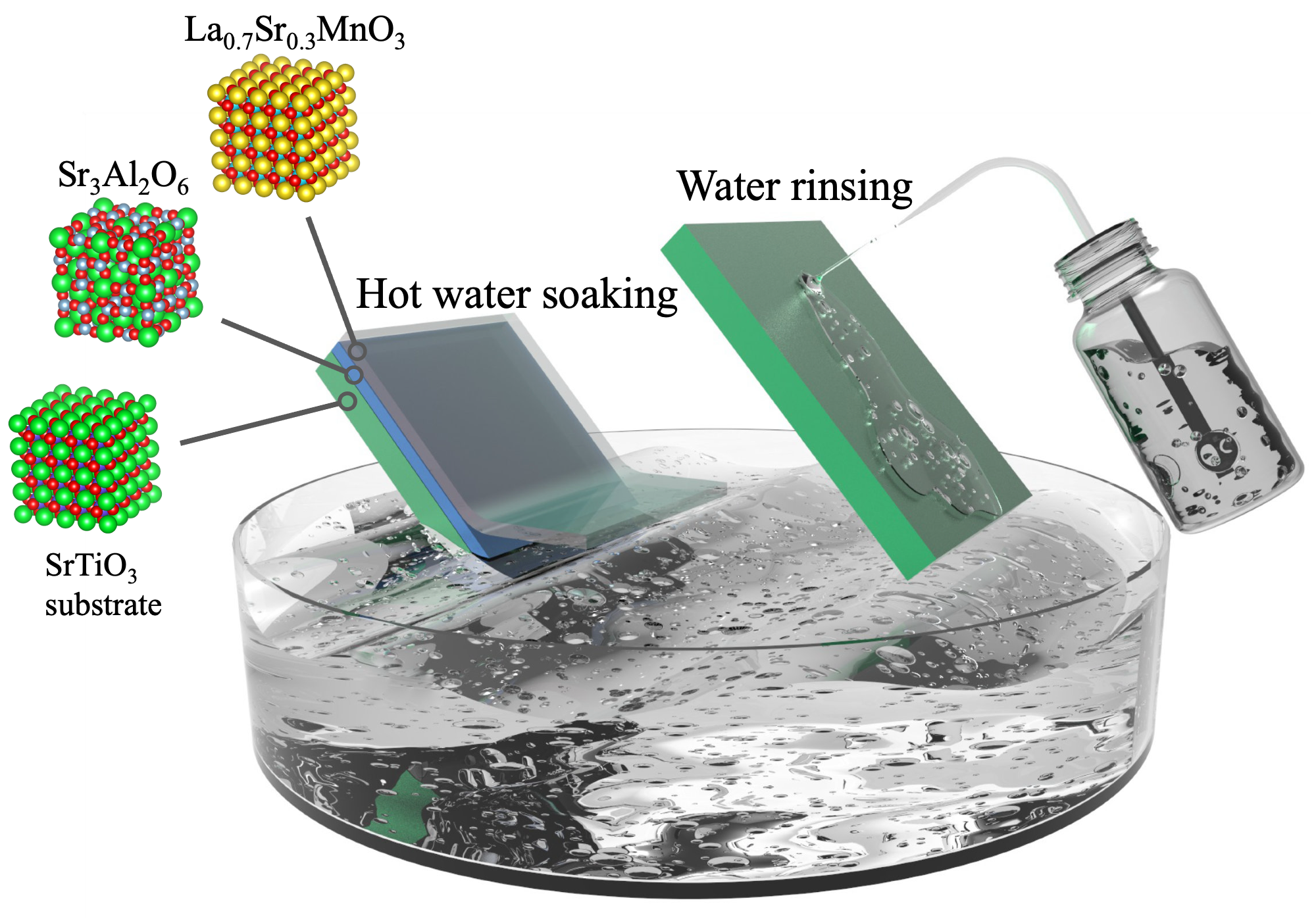}
	\caption{The schematic substrate recycling procedure in preparing high-quality functional complex oxide thin films.}
	\label{Fig1}
\end{figure}

\begin{figure*}[ht]
\centering
	\includegraphics[width=\linewidth]{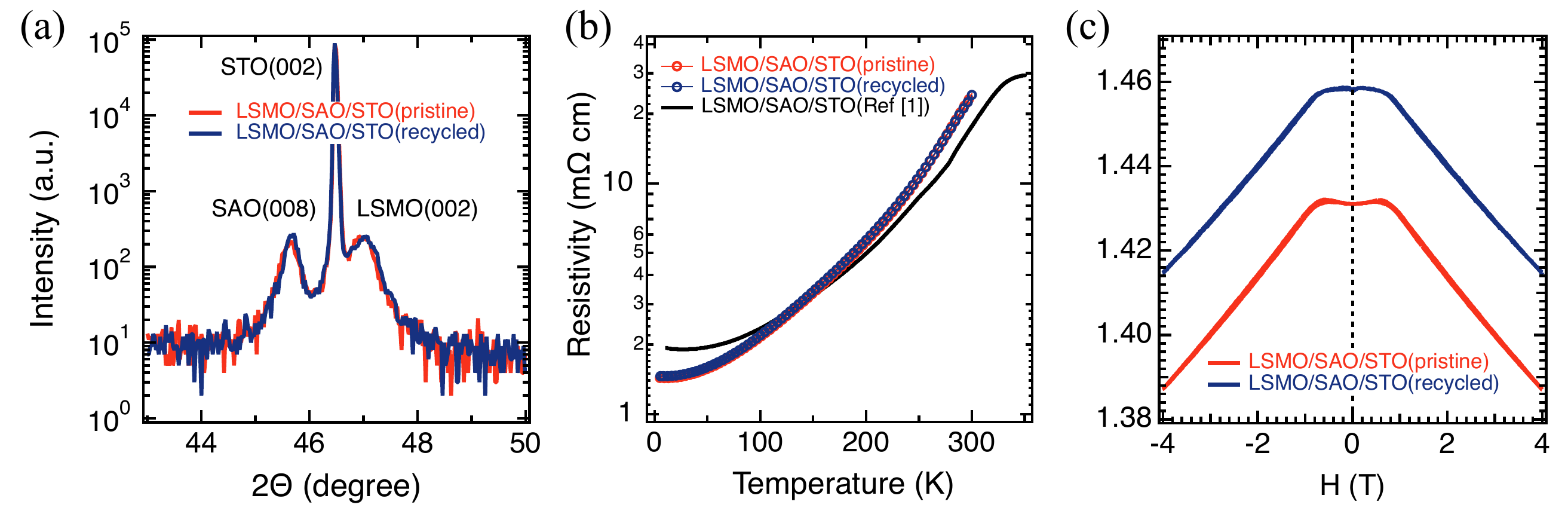}
	\caption{(a) XRD patterns of LSMO/SAO/STO(pristine) and LSMO/SAO/STO(recycled). The resistivity as a function of temperature (b) and magnetic field (c) of the fresh growth and regrowth samples. The reference black curve in (b) was reproduced from Ref. \cite{Lu2016}.}
	\label{Fig2}
	\end{figure*}

\section{Results and Discussions}

The epitaxial lift-off using an SAO sacrificial layer and sequential substrate recycling processes are illustrated in Figure \ref{Fig1}, details of the experiments can be found in Section \ref{sec3}. The SAO buffer layer and a representative functional LSMO layer were grown on STO(001) substrates by pulsed laser deposition (PLD). Figure \ref{Fig2} shows the good reproduction of pristine and recycled SAO and LSMO layers detected by XRD and transport properties. XRD patterns for LSMO/SAO thin films on the pristine and recycled STO (MTI Corporation) are displayed in Figure \ref{Fig2}(a). As seen, The (008)SAO and (002)LSMO peaks are located on the left and right side of (002)STO peak, consistent with the previous study \cite{Lu2016}. No detectable impurity phases were detected in a wider range (20--80\degree) XRD pattern (not shown here). The almost identical XRD patterns indicate that the LSMO/SAO grown on the recycled STO possessed a non-degraded crystallinity as compared to that grown on the pristine STO. Thus, the recycled substrate should have returned to its original state. 

\begin{figure}[h!]
\centering
	\includegraphics[width=\linewidth]{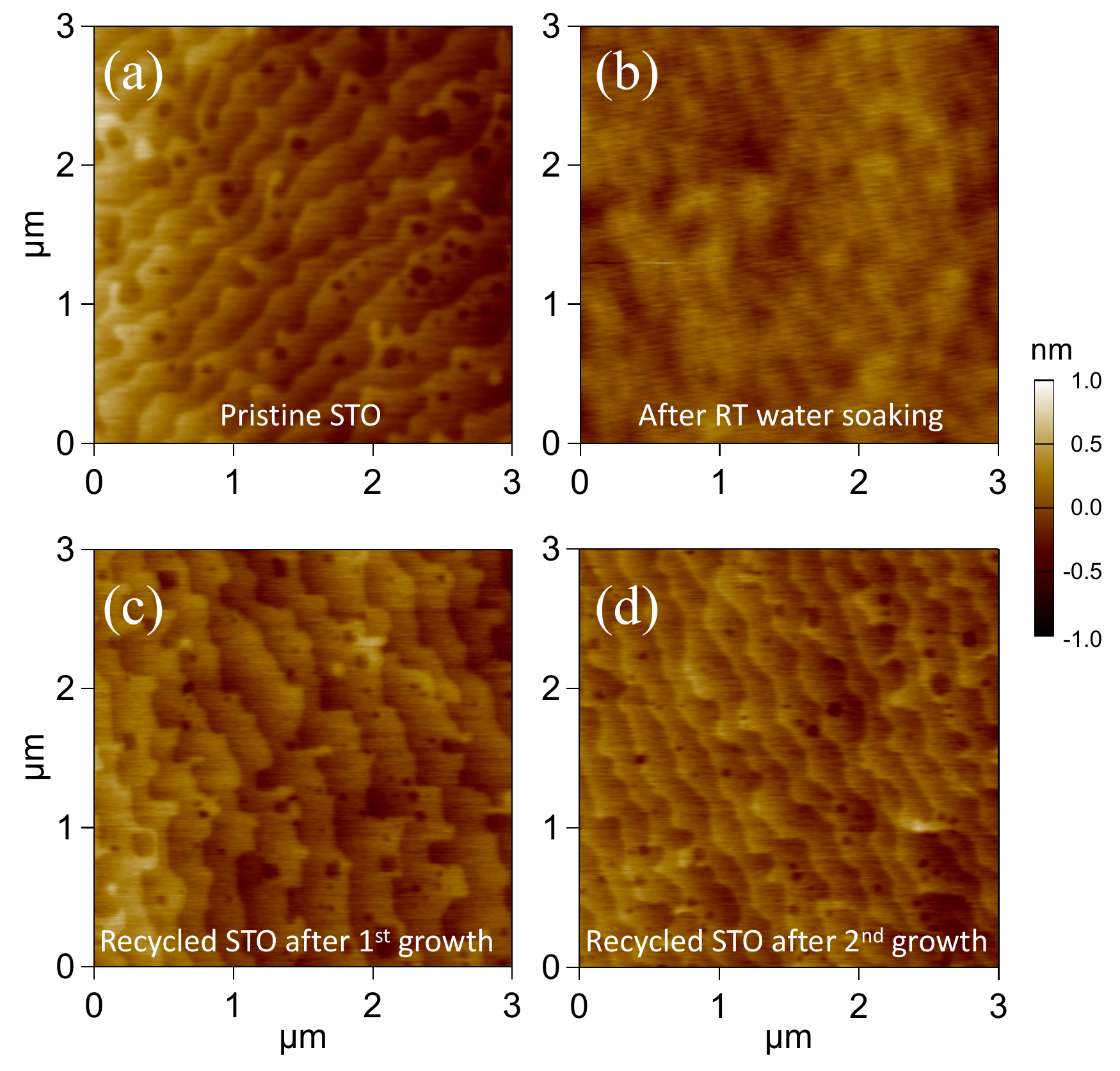}
	\caption{AFM topographies of the pristine STO substrate (a), substrate after room-temperature water soaking (b), recycled substrate by 80 \degree C hot water treatment after first time growth (c), and recycled substrate after second time growth (d), respectively.}
	\label{Fig3}
\end{figure}
\begin{figure*}[t!]
\centering
	\includegraphics[width=\linewidth]{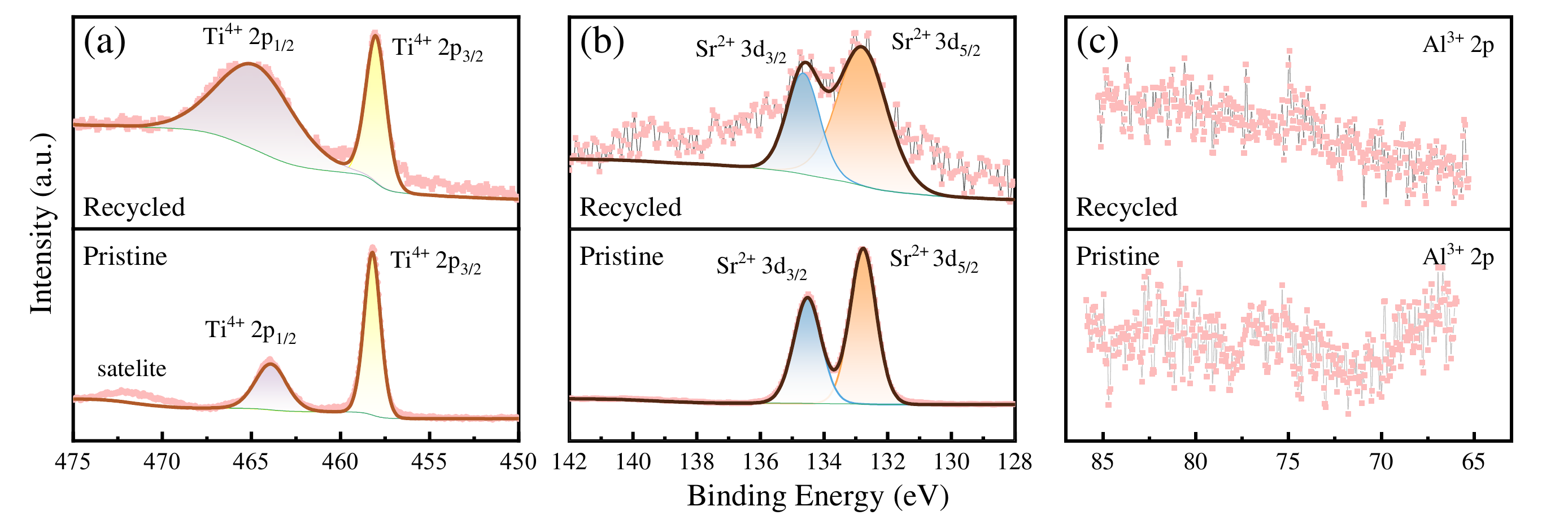}
	\caption{XPS of the pristine (lower panel) and recycled (upper panel) STO substrates, (a) Ti 2p, (b) Sr 3d, (c) Al 2p.}
	\label{Fig4}
\end{figure*}

To further verify the quality of the recycled LSMO/SAO/STO heterostructures compared with pristine one, we measured the temperature- and magnetic field-dependent resistivity of aforementioned two samples by a physical property measurement system (PPMS, Quantum Design). As seen in Figure \ref{Fig2}(b), the resistivity versus temperature ($R$--$T$) curves of both samples perfectly coincided. The black curve in Figure \ref{Fig2}(b) was a reproduction from a very similar heterostructure of LSMO/SAO/STO from Ref. \cite{Lu2016}, it was just this seminal paper which originally reported the SAO as a sacrificial layer to fabricate freestanding perovskite oxides. Indeed, we see that the $R$--$T$ of our samples is comparable to, or slightly better than that in Ref. \cite{Lu2016}, since at very low temperature (e.g. $\leqslant$ 20 K), the Anderson weak localization (upturn of resistivity at low temperature) was not present. In addition, we also detected the ferromagnetic properties by field-dependent resistivity (\textit{MR}) curves. As shown in Figure \ref{Fig2}(c), these two \textit{MR} curves at 5 K are quite similar in lineshape, with approximately identical coercive field and \textit{MR} value. The \textit{MR} data of our samples is very similar to a very conventional lineshape of LSMO \cite{Liao2014}. Combining the XRD patterns and transport properties, we conclude that the functional layer, as exemplified by LSMO, can be impeccably fabricated on a recycled substrate with well reproducible properties. We have also carried out experiments on the effect of environmental moisture on film stability (Figure S1, Supporting Information), which indicated that sample surface topography deteriorated in a high relative humidity (RH, here 99\%) at 25 \degree C, while it remained unchanged in a low relative humidity (here, 20\%).

To validate the reproducibility and recoverability of recycled substrate in our process, here, we utilized AFM (Asylum Research) equipped with an FM-LC tip (Adama Innovations) to detect the topographies of pristine and recycled substrates \cite{Lu2016}. As displayed in Figure \ref{Fig3}(a), the pristine substrate shows a standard step-and-terrace surface structure as expected. After the lift-off process with room-temperature water, some residues adhered to the sample surface with the absence of the step-and-terrace surface of the substrate, as shown in Figure \ref{Fig3}(b). We ascribed it to the colloid formed by the reaction of aluminate with water. Remarkably, after the substrates were treated with hot water, both recycled substrates after first and second-time growths, have restored to the step-and-terrace surfaces without any observable deterioration, as shown in Figure \ref{Fig3}(c)--(d). This is consistent with the similar reflection high energy electron diffraction (RHEED) patterns of pristine and recycled substrates (Figure S2, Supporting Information), which indicates that the surface structure of substrate is unaffected by the adequate hot water soaking. Note that, the step edges of the substrates are not perfectly ordered due to the slightly poor pristine substrate quality. In addition, different growth temperatures were used to prepare LSMO/SAO thin films, which did not affect the surface topography of the recycled substrates (Figure S3, Supporting Information). Nevertheless, this undoubtedly manifests that the substrates were successfully recycled by hot water treatment. 

Finally, the surface chemical states of the pristine and recycled STO substrates were examined by XPS as shown in Figure \ref{Fig4}. XPS is a surface-sensitive technique, which is one of the best methods to detect residues at trace levels. As expected, similar Ti 2p and Sr 3d peaks were seen in pristine and recycled STO substrates, see Figure \ref{Fig4}(a)--(b) \cite{Vasquez1992}. However, no Al$^{3+}$ 2p peak from residual adjacent Sr$_3$Al$_2$O$_6$ layer was identified (Figure \ref{Fig4}(c)), which is consistent with our AFM characterizations and reveals the atomic-level recycling of STO substrates.

\section{Conclusion}

In summary, our findings corroborate that the substrates used in fabricating freestanding functional complex oxides by the sacrificial SAO method can be non-destructively reused, and the recycled functional layer is almost identical to pristine state. The recycling procedure is exceptionally facile and straightforward, i.e., adequate hot water soaking and rinsing are capable to refresh the surface of the substrates. This technique may reduce the cost of research activities in this field, especially fabrication of large-scale flexible devices with freestanding complex oxides. In addition, by artificially including a degradable buffer SAO layer, such substrate recycling can be popularized even in non-freestanding complex oxide synthesis. We also highly advocate the researchers to scrutinize whether an analogous recycling technique can be developed in other epitaxial lift-off processes.

\section{Experimental Section} \label{sec3}

\textbf{Films Growth and Structural Characterization}\\
The SAO buffer layer and a representative functional LSMO layer were grown on STO(001) substrates by pulsed laser deposition (PLD). Prior to the growth, the STO was treated with a mixture of concentrated nitric and hydrochloric acids according to the protocol used in Ref. \cite{Kareev2008}. The growth condition for SAO was 720 \degree C (measured by a pyrometer) in $10^{-6}$ Torr oxygen with a KrF excimer laser (energy density of $\thicksim$ 1.5 J/cm$^2$, repetition rate of 2 Hz), while the LSMO was grown in 200 mTorr oxygen with all other parameters same to those of SAO. X-ray diffraction (XRD, Malvern Panalytical) measurements were used to verify the crystal structure, phase purity, and orientation of these thin films.

\textbf{Substrate Refreshment and Surface Topography}\\
After the growth, the sample was cut into two pieces. One out of them was soaked into room-temperature deionized water for $\thicksim$ 6 hours to acquire the freestanding LSMO. Thereafter, we heated the water to $\thicksim$ 80 \degree C for $\thicksim$ 6 hours to get rid of the tiny amounts of debris of SAO on the surface of STO (The pH values of the deionized water were 6.94 and 6.22 at 20 \degree and 80 \degree, respectively.). The surface Topography was detected by atomic force microscopy (AFM, Asylum Research) equipped with an FM-LC tip (Adama Innovations). After sequential rinsing by deionized water, the exact same growth conditions were utilized to repeatedly fabricate LSMO/SAO thin films on this recycled STO substrate.

\textbf{Electrical Transport}\\
The temperature dependence of the resistivity and magnetoresistance was carried out by Physical Property Measurement System (PPMS, Quantum Design DynaCool system) via  the Vander Pauw method.

\textbf{X-ray Photoelectron Spectroscopy}\\
X-ray Photoelectron Spectroscopy was performed by ESCALAB 250Xi X-ray Photoelectron Spectrometer (Thermo Fisher Scientific) equipped  with  a  monochromatic  Al  $k_\alpha$ X-ray  source. High-resolution  spectra  were  acquired  for  surface  elemental identification, which was fitted based on Lorentzian functions broadened by a Gaussian function.

\medskip
\textbf{Acknowledgements}

L. W. acknowledges supports from National Natural Science Foundation of
China (Grant No. 52102131) and Yunnan Fundamental Research Projects (Grant No. 202101BE070001-012 and 202201AT070171).

\bibliographystyle{apsrev4-2}
\bibliography{ref}

\end{document}